# Graphene doping to enhance flux pinning and supercurrent carrying ability in magnesium diboride superconductor

X. Xu, S. X. Dou,[a] X. L. Wang, J. H. Kim
*Institute for Superconducting & Electronic Materials, University of Wollongong, Wollongong, NSW, 2522, Australia*

J. A. Stride, M. Choucair
*School of Chemistry, University of New South Wales, Sydney, NSW 2052, Australia*
*Australian Nuclear Science and Technology Organisation, PMB 1, Menai, NSW 2234, Australia*

W. K. Yeoh, R. K. Zheng, S. P. Ringer
*Electron Microscope Unit, University of Sydney, NSW 2052, Australia*

It has been shown that graphene doping is sufficient to lead to an improvement in the critical current density - field performance ($J_c(B)$), with little change in the transition temperature in $MgB_2$. At 3.7 at% graphene doping of $MgB_2$ an optimal enhancement in $J_c(B)$ was reached by a factor of 30 at 5 K and 10 T, compared to the un-doped sample. The results suggested that effective carbon substitutions by grapheme, 2D nature of grapheme and the strain effect induced by difference thermal coefficient between single grapheme sheet and $MgB_2$ superconductor may play an important role in flux pinning enhancement.

Correspondence and requests for materials should be addressed to S. X. Dou (shi_dou@uow.edu.au)

Substitutional chemistry can modify, in a controlled way, the electronic structures of superconductors and their superconducting properties, such as the transition temperature ($T_c$), critical current density ($J_c$), upper critical field ($H_{c2}$), and irreversibility field ($H_{irr}$). In particular, carbon containing dopants, including nano-meter sized carbon (nano-C), silicon carbide (SiC), carbon nanotubes (CNTs), hydrocarbons/carbohydrates, and graphite are effective means to enhance the $J_c$- field dependence and $H_{c2}$[1-11]. In this work, it will be seen the graphene as one kind of caborn source dopant, how it is useful to incorporation into $MgB_2$ and it is expected that $H_{c2}$ and the flux pinning properties should be improved. However, until now there has been no report on the effects of graphene doping on the superconductivity of $MgB_2$, partly due to the unavailability of graphene on a suitable scale. Recently, high-throughput solution processing of large-scale graphene has been reported by a number of groups[12-17].

Based on the works of Choucair *et al.*[18], sufficient quantities of graphene were obtained for doping the bulk $MgB_2$ samples via a diffusion process. The crystalline Boron powder (0.2 to 2.4 µm) 99.999% without and with graphene[18] was prepared by ball milling with toluene medium. Then the powders were dried in a vacuum oven to evaporate the medium. These powders were mixed and pressed into pellets. The pellets were then put into an iron tube filled with Mg powder (-325mesh 99%). The samples were sintered at 850°C for 10 hrs in a quartz tube; the heating rate was $5^oCmin^{-1}$ under high purity argon (Ar 99.9%) gas. The phase and crystal structure of all the samples were investigated by X-ray diffraction (XRD). $T_c$ was defined as the onset temperature at which diamagnetic properties were observed. The magnetic $J_c$ was derived from the width of the magnetization loop using Bean's model by a Physical Properties Measurement System (PPMS). Transport measurements for resistivity ($\rho$) were done using a standard AC four probe method. In addition, $H_{c2}(T)$ and $H_{irr}(T)$ were defined as the fields where the temperature dependent resistance at constant magnetic field $R(H_{c2}, T) = 0.9R_{ns}$ and $R(H_{irr}, T) = 0.1R_{ns}$ with $R_{ns}$ being the normal state resistance near 40 K.

The common format $Mg(B_{1-x}C_x)$, x=0, 0.037, and 0.087 was used. The composition of graphene doped $MgB_2$ were 0, 3.7, and 8.7 at%, and as such, the sample names are designated as G000, G037, and G087, respectively. We demonstrate that the graphene doping can result a greater enhancement of the critical current density ($J_c$) by over one order of magnitude in high magnetic fields. The $J_c$ achieved is as high as $2.0 \times 10^4$ Acm$^{-2}$ at 5 K and 8 T magnetic field for a graphene dopant level of only 3.7 at%, with only a slight corresponding drop in $T_c$. This improvement is likely caused by the following factors: the single carbon sheet with two dimensional (2-D) geometry, the negative thermal expansion coefficient, and high thermal conductivity. However, further studies are needed to clarify the dominating mechanisms responsible for the enhancement.

The lattice parameters, *a*, *c*, the ratio of *a/c,* grain size, strain, and full width at half maximum of the representative peak (110) calculated from the XRD patterns are shown in Table 1. Both the *a*-axis and *c*-axis parameters vary little with increasing graphene doping level of 3.7%, apart from G087 sample, which shows a notable decrease in the *a*-axis parameter, suggesting that carbon (C) is likely to be partially substituted into the boron (B) sites, leading to a slight drop in $T_c$ (36.7 K) for the G087 sample. We also observed that the full width at half maximum (FWHM) of the (110) peak increases gradually with increasing graphene dopant level. Such a peak broadening is caused by both grain size reduction and an increase in lattice strain. The calculated results on grain size and lattice strain from a Williamson-Hall plot[18] are given in Table 1. Also, the $T_c$ onset determined from the AC susceptibility measurement is 38.9 K for the un-doped sample, dropping only slightly to 37.7 and 36.7 K for the G037 and G087 samples, respectively.

The temperature dependence of the resistivity ($\rho$) measured in different fields is shown in Fig. 1. The graphene doped samples have higher resistivity than the un-doped MgB$_2$ sample, indicating that electron scattering occurs at higher doping levels to some extent. However, it should be pointed out that the increase in resistivity is much smaller than for any other forms of carbon doped MgB$_2$[1-9]. The $J_c$ is actually greatly enhanced in high magnetic fields for the graphene doped samples due to enhanced flux pinning.

Fig. 2 shows the $J_c(B)$ curves at 5 K and 20K for all the samples, which were sintered at 850°C for 10 hours. The $J_c(B)$ values for all the doped samples are higher than the un-doped sample at high fields. The sample G037 gives the highest $J_c$ at high fields: $J_c$ increases by a factor of 30 at 5 K for the field of 10 T, as compared to the un-doped sample, G000. Even though the $J_c$ in the low field regime is depressed, a higher doping level (G087), still results in the rate of $J_c$ dropping much slower than the un-doped sample, clearly indicating strong flux pinning induced by the graphene doping. The most significant effect of graphene doping is the high effectiveness of graphene to improve flux pinning at lower doping levels, which distinguishes graphene from any other C containing dopants, for example, the $J_c$ for G037 reached 20,000 A/cm$^2$ at 5 K and 8 T, exceeding or matching the best $J_c$ resulting from dopants such as SiC, CNT, and carbohydrates at their optimal doping level of 10 at%[1,2,5-8], as well as nano-C at its optimal doping level of 5-6.4 at.%[3,4,9,10]. In the latter case, $T_c$ is noticeably reduced to temperatures as low as 30 K. Compared to the graphite doped MgB2 pallets prepared through the ball-milling and HIP the $J_c$ of graphite doping is better than graphene doping at 5 K[11], but at 20K, the graphene doping is much better than graphite, for example, the $J_c$ is more than 500 A/cm$^2$ at 6T for graphene doped while only 10 A/cm$^2$ for graphite doped $MgB_2$[11]. In comparison, low levels of graphene doping have little effect on $T_c$ and cause only a very small increase in impurities, not compromising the significant enhancement in $J_c$ in high fields by the degradation in low-field $J_c$, which is a common issue for all other C based dopants. In order to see the difference with other C based dopant, the same preparation route was applied to 5 at% nano-C doped sample and the resultant decrease in $J_c$ at 20K can be seen in the Fig. 2, This is because the $T_c$ is only 34K for this sample.

Fig. 3 shows the upper critical field, $H_{c2}$, and the irreversibility field, $H_{irr}$, versus the normalised $T_c$ for all the samples. It should be noted that both $H_{c2}$ and $H_{irr}$ are increased by graphene doping. This indicates that graphene doping gives a much stronger improvement in flux pinning than in the upper critical field. The latter is closely related to the carbon substituting for B.

The mechanism for enhancement of $J_c$, $H_{irr}$, and $H_{c2}$ by carbon containing dopants has been well studied. The C can enter the $MgB_2$ structure by substituting into B sites, and thus $J_c$ and $H_{c2}$ are significantly enhanced due to the increased impurity scattering in the two-band $MgB_2$[20]. Above all, C substitution induces highly localised fluctuations in the structure and $T_c$, which have also been seen to be responsible for the enhancements in $J_c$, $H_{irr}$, and $H_{c2}$ by SiC doping[1]. Furthermore, residual thermal strain in the $MgB_2$-dopant composites can also contribute to the improvement in flux pinning[21]. In the present work, the C substitution for B (up to 3.7 at.%) graphene doping is lower, from the table 1, the change of the a-parameter is too smaller, according to Avdeev *et al* result[22], the level of C substitution, x in the formula $Mg(B_{1-x}C_x)$, can be estimated as $x=7.5 \times \Delta(c/a)$, where $\Delta(c/a)$ is the change in c/a compared to a pure sample. If there is any at all - as both the *a*-axis and the *c*-axis lattice parameters determined from the XRD data showed little change within this doping range. This is in good agreement with the small reduction in $T_c$ over this doping regime. At 8.7 at% doping, there is a noticeable drop in the *a*-axis parameter, suggesting C substitution for B, which is also consistent with the reduction in $T_c$. The source of C could be the edges of the graphene sheets, although the graphene is very stable at the sintering temperature (850°C), as there have been reports of graphene formation on substrates at temperatures ranging from 870-1320°C[23].

The significant enhancement in $J_c$ and $H_{irr}$ for G037 can not be explained by C substitution only. The strict 2-D nature of graphene in irregular shapes may make the sheets strong pinning centres. It is also important to note that the coefficient of thermal expansion for graphene is negative[24,25] up to 2300 K, while for $MgB_2$, this is not only very large, but highly anisotropic, with a larger coefficient of thermal expansion in the *c*-axis direction from room temperature to 900°C[26]. Upon cooling, $MgB_2$ is subjected to high tensile strain, a consequence of the graphene expanding in accordance with the negative coefficient of thermal expansion. As a result, the large strain along the graphene sheet will induce defects in $MgB_2$, contributing to the enhancement in flux pinning and $H_{irr}$. The strain induced enhancement of flux pinning was also observed in coated superconductors[27] due to lattice mismatch and in SiC-$MgB_2$ composites as a result of residual thermal strain[21]. In the case of SiC-$MgB_2$ composite, the coefficient of thermal expansion for SiC is not negative, but is smaller than that of $MgB_2$. Whilst graphene has a negative coefficient of thermal expansion and hence a larger effect on strain, a larger extent of defects is expected in graphene doped $MgB_2$.

The microstructure revealed by high resolution transmission electron microscope (TEM) observations show that G037 sample has grain size of 100-200 nm which is consistent with value of the calculated grain size in table 1. The graphene doped samples have relatively higher density of defects compared with the undoped sample under TEM images as shown in figure 4(a) and (c). The density of such defects is estimated to be 1/3 areas of TEM images, indicating high density in the doped samples. In Figures 4(b) it should be noted that the order of fringes varies from grain to grain, indicates that the defect is due to highly anisotropic of the interface. Similar fringes have been reported in the MgB$_2$[21] where these fringes were induced by tensile stress with dislocations and distortions which were commonly observed in the areas. As the graphene doped samples were sintered at 850$^o$C for 10 hrs, the samples are expected to be relatively crystalline and contain few defects. The large amount of defects and amorphous phases on the nanoscale can also be attributed to the residual thermal strain between the graphene and the MgB$_2$ after cooling. Defects on the order of the coherence length of the beam, $\xi$, can also play a role as effective pinning centres that are responsible for the enhanced flux pinning and $J_c$ in the graphene doped MgB$_2$.

In summary, the effects of graphene doping on the lattice parameters, $T_c$, $J_c$, and flux pinning in MgB$_2$ were investigated over a range of low doping levels. It was found that substitution of C for B enhances the flux pinning slightly depressing $T_c$. By controlling the processing parameters, an optimised $J_c(B)$ performance is achieved at a doping level of 3.7 at.%. Under these conditions, $J_c$ was enhanced by an order of magnitude at 8 T and 5 K. The combination of the 2-D geometry of graphene, low C substitution for B, and in particular, residual thermal strain effects between graphene and MgB$_2$ are proposed to be responsible for the enhancement of flux pinning in high fields. The strong enhancement of $J_c$, $H_{c2}$, and $H_{irr}$ with low levels of graphene doping is promising for large-scale MgB$_2$ wire applications.

**Table I**: The full width at half maximum (FWHM) of the (110) peak, the lattice parameters, and the transition temperature ($T_c$) for the MgB$_2$ samples, made with 0, 3.7, and 8.7 at% graphene doping via a diffusion process.

| Sample | Lattice Constants | | | Grain Size (nm) | Strain (%) | FWHM (110) (°) | $T_c$ (onset) (K) |
|---|---|---|---|---|---|---|---|
| | *a* (Å) | *c* (Å) | *c/a* | | | | |
| G000 | 3.084(1) | 3.525(1) | 1.143(1) | 216(10) | 0.1198(188) | 0.288 | 38.9 |
| G037 | 3.082(1) | 3.527(1) | 1.144(1) | 170(8) | 0.1685(250) | 0.400 | 37.7 |
| G087 | 3.075(1) | 3.525(1) | 1.146(1) | 171(11) | 0.1782(330) | 0.414 | 36.7 |

**Figure Captions:**

**Figure 1**: The temperature dependence of the resistivity (ρ) measured in different fields for doped and undoped samples.

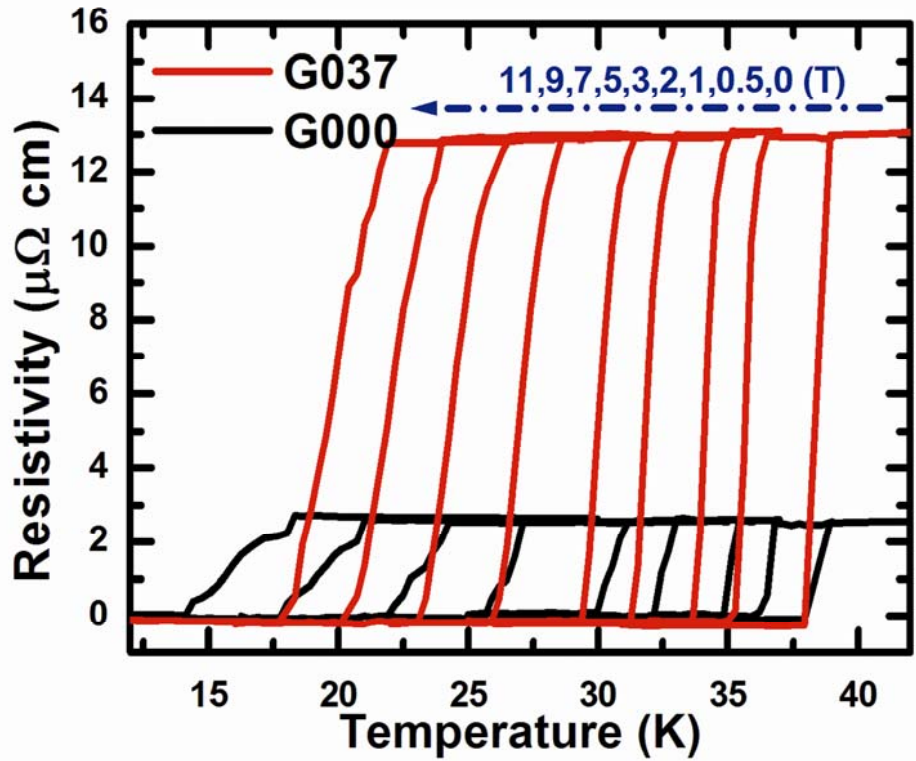

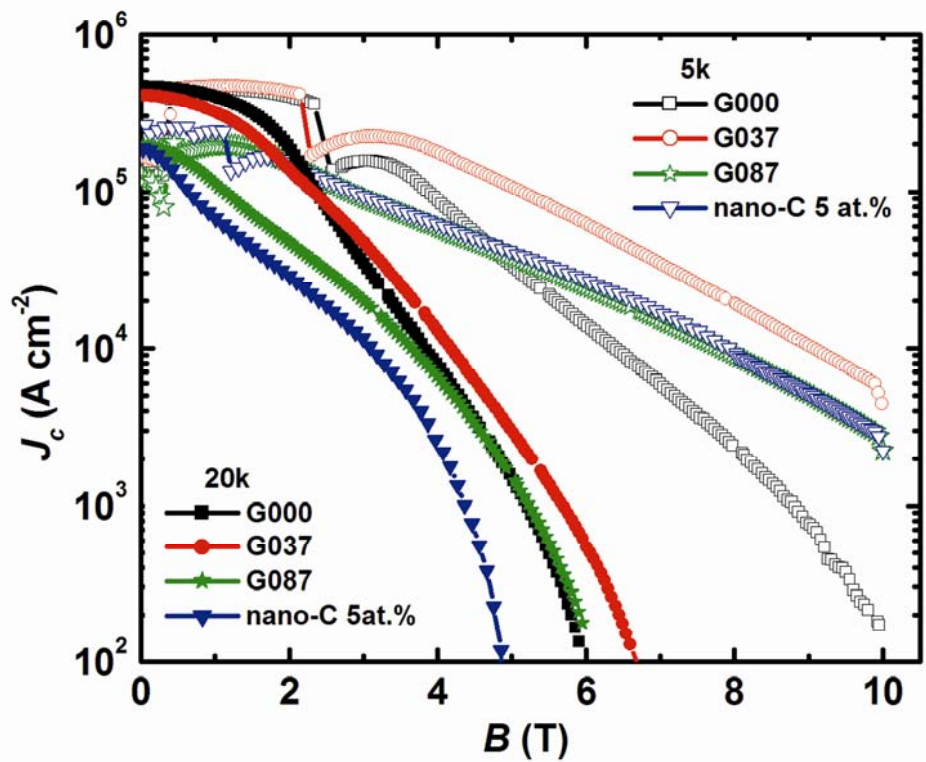

**Figure 2**: Critical current density as a function of magnetic field at 5 K and 20K for with and without graphene doped samples. 5 at% nano-C doped sample for a comparable result at the same sample preparation route.

**Figure 3**: Upper critical field, $H_{c2}$, and irreversibility field, $H_{irr}$, versus normalised transition temperature, $T_c$, for all graphene doped and undoped MgB$_2$ samples.

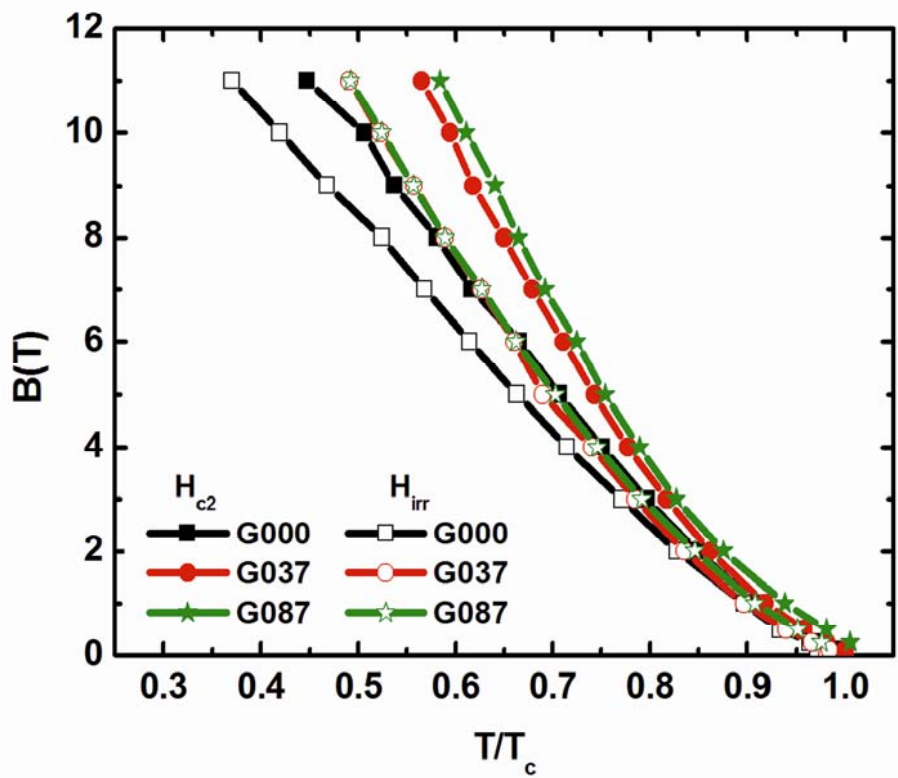

**Figure 4**: (a) TEM image showing the defects with grains of the G037 sample with order of fringes varies between grains. Defects and fringes are indicated by arrow, and (b) HRTEM image of fringes. TEM images show large amount of defects and fringes can be observed in the graphene doped sample G037. (c) TEM image of the undoped sample for reference.

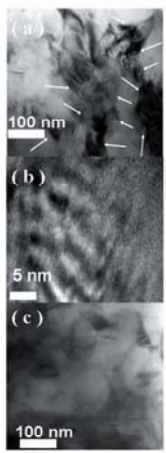